\documentclass[journal=jpclcd,manuscript=letter]{achemso}

\usepackage{graphicx}
\usepackage{bm}
\usepackage{mathptmx}
\usepackage{epstopdf}
\usepackage{booktabs}
\usepackage{multirow}
\usepackage{hhline}
\usepackage{dcolumn}
\usepackage{subfigure}
\usepackage{amsmath}
\usepackage{float}
\usepackage{chemformula}
\usepackage[version=3]{mhchem}
\usepackage{chemfig}
\usepackage{soul}
 
\title
{Insights into Heterogeneous Catalysis on Surfaces with 3d Transition Metals: Spin-Dependent Chemisorption Models and Magnetic Field Effects}

\author{Satadeep Bhattacharjee}
\email{s.bhattacharjee@ikst.res.in}
\affiliation{Indo-Korea Science and Technology Center (IKST),  Bangalore-560064, India}
\author{Swetarekha Ram}
\affiliation{Indo-Korea Science and Technology Center (IKST),  Bangalore-560064, India}
\author{Seung-Cheol Lee}
\affiliation{Electronic Materials Research Center, Korea Institute of Science and Technology, Seoul 02792, Republic of Korea}
\email{leesc@kist.re.kr}

\begin{tocentry}
	\vspace{1cm}
	\centering
	\includegraphics[width=\textwidth]{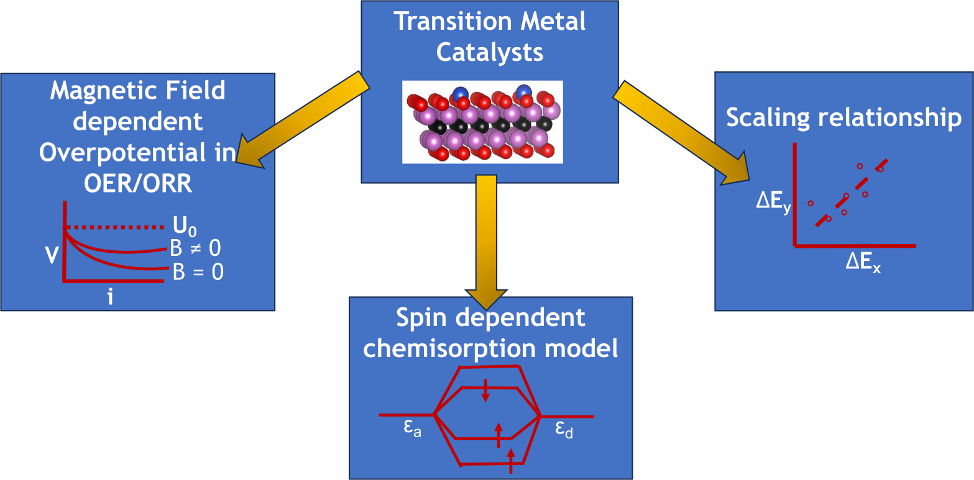}
\end{tocentry}

\begin{document}
\begin{abstract}
This perspective article provides an overview of recent developments in the field of 3d transition metal (TM) catalysts for different reactions including oxygen-based reactions such as Oxygen Reduction Reaction (ORR) and Oxygen Evolution Reaction (OER). The spin moments of 3d TMs can be exploited to influence chemical reactions, and recent advances in this area, including the theory of chemisorption based on spin-dependent d-band centers and magnetic field effects, are discussed. The article also explores the use of scaling relationships and surface magnetic moments in catalyst design, as well as the effect of magnetism on chemisorption and vice versa. In addition, recent studies on the influence of a magnetic field on the ORR and OER are presented, demonstrating the potential of ferromagnetic catalysts to enhance these reactions through spin polarization.
\end{abstract}
\maketitle

Heavy nonmagnetic metals have traditionally been utilized as catalysts and are not suitable for investigating the role of magnetism in heterogeneous catalysis. However, in recent years, there has been a shift in emphasis towards using simple 3d-transition metals (TMs) as catalysts~\cite{3d1,3d2,3d3,3d4,Emily,miao,gracia,liu2023recent}. Because of their abundance and cheap cost, these metals are frequently used as alloying elements in binary or ternary alloys-catalysts. 3d transition metals are frequently alloyed with heavy metals in electrochemical reactions such as ORR and OER~\cite{3d5,3d6}. Most importantly, 3d TMs bring a new parameter to control or manipulate the chemical reactions: \textit{The spin moments}.
In recent years, magnetism has been shown to play a significant role in heterogeneous catalysis, particularly in the use of magnetic nanoparticles and magnetic fields. Magnetic nanoparticles show promising applications in heterogeneous catalysis due to their ease of separation and good reusability~\cite{nano}. Mtangi \textit{et al} showed that spin-selectivity eliminates the formation of hydrogen peroxide, which is a common problem in water splitting, while increasing the overall current through the cell. 
Magnetic fields can influence the energy levels of active species by interacting with their spin states, which can affect catalytic properties and alter the selectivity of the reaction~\cite{multiscale}. 
In the case of alloy catalysts, the effect of magnetism becomes even more prominent due to the charge transfer between the components. The interplay between charge transfer and magnetism can affect the activity and selectivity of alloy catalysts, and this has been investigated in several studies~\cite{CT1,CT2}. 
It was highlighted that the angular momentum plays a crucial role in reactions like OER when one of the reactants or reaction intermediate is magnetic~\cite{torun2013role}. In non-magnetic anodes, oxygen can only be produced in an excited non-magnetic state without violating the conservation of angular momentum. This is because neither water nor hydrogen are magnetic. The two lowest excited states of the oxygen molecule are singlet states, 1 and 1.6 eV above the triplet ground state. It is believed that for nonmagnetic anodes the high overpotential is related to the notion that oxygen is first generated in its nonmagnetic excited state and slowly decays to the ground state by higher order processes~\cite{torun2013role}. In such scenarios, including the effects of surface magnetism in the OER could be of great relevance. The ruthenium dioxide (110) surface is magnetic, allowing the production of ground-state oxygen while conserving angular momentum. \textit{Ab initio} calculations of the electronic structure show that this surface itself carries magnetic moments. The scenarios mentioned above also indicate that the magnetic field could manipulate mechanisms such as OER or ORR.
The role of conservation of spin angular momentum in real electron transfer during OER or ORR was further discussed in a recent work~\cite{gracia2019itinerant} where the authors point out that conservation of angular momentum ensures that electron transfer occurs between overlapping orbitals of reactants and catalysts, following the Goodenough-Kanamori rules.

From the theoretical point of view in heterogeneous catalysis the trend of chemical reactivity is often explained in terms of the so-called d-band center model~\cite{d-band1,d-band2}. The d-band center model has been widely employed in heterogeneous catalysis, providing fundamental insights into the link
between catalyst electronic structure and reactivity and selectivity. The model has been applied to a wide range of catalytic systems, including transition metal catalysts and other types of catalysts, and has been used to understand the effect of different factors on the reactivity and selectivity of the catalysts, such as their composition, structure, and coordination environment.

The concept of the d-band center model was borrowed from a much older and original model of chemisorption, namely the Newns-Anderson model, which described the interaction of an adsorbate level $\varepsilon_a$ in a state $|a>$ with a continuous set of states $|k>$ via the Hamiltonian is given by~\cite{NW},
\begin{equation}
\hat{H}=\varepsilon_{a}c_{a}^{+}c_{a}+\sum_{k}\varepsilon_{k}c_{k}^{+}c_{k}+\sum_{k}(V_{a k}c_{k}^{+}c_{a}+V_{a k}^{*}c_{a}^{+}c_{k}),
\label{Newns-Anderson}
\end{equation}
Where the first term refers to the adsorbate energy and the second term describes the metal states. The third term describes the coupling between the two ($V_{a k}=<a|\hat{H}| k>$). The above Hamiltonian can be solved using  Green's function approach. Introducing the Green's function, $G(\varepsilon)=\sum_m \frac{|m><m|}{\varepsilon-\varepsilon_m+i \delta}$, we obtain the chemisorption energy as,
\begin{equation}
{\Delta E}=2\big({\frac{1}{\pi}}\int^{E_F}_{-\infty} Arc~tan\frac{\Delta(\varepsilon)}{\varepsilon-\varepsilon_{a}-\Lambda(\varepsilon)}d\varepsilon-\varepsilon_a \big)
\label{sol1}
\end{equation}
Here $\left|m>=c_{a_m}\right| a>+\sum_k c_{k_m} \mid k>$ is the $m^{th}$ eigenstate of the Hamiltonian $\hat{H}$.
In the above equation, $\Delta\left(\varepsilon\right)=\pi\sum_{k}\vert V_{a k}\vert^{2}\delta\left(\varepsilon-\varepsilon_{k}\right)$ is the imaginary part of the self-energy or chemisorption function while $\Lambda\left(\varepsilon\right)=\frac{P}{\pi}\int\frac{\Delta\left(\varepsilon^{\prime}\right)}{\varepsilon-\varepsilon^{\prime}}d\varepsilon^{\prime}$ is the corresponding real part. $E_F$ is the Fermi energy.

The d-band center model proposed by Hammer and Norkskov is similar to the solution of the Newns-Anderson model within the narrow-band limit where the entire d-band is replaced by a single energy level known as d-band center, $\varepsilon_{\mathrm{d}}=\frac{\int_{-x}^{\infty}n_{\mathrm{d}}(\varepsilon)\varepsilon~d\varepsilon}{\int_{-x}^{\infty}n_{\mathrm{d}}(\varepsilon)d\varepsilon}$. Where $n_{\mathrm{d}}(\varepsilon)$ is the density of states of the metal d-states. The chemisorption function then becomes a delta function at $\varepsilon=\varepsilon_{\mathrm{d}}$. If we consider the metal adsorbate coupling, $V_{ak}\sim V$ is independent of $k$, then  $\Delta=\pi V^2\delta\left(\varepsilon-\varepsilon_{d}\right)$ is constant. The real part of the self-energy turns to be $\Lambda\left(\varepsilon\right)=\frac{V^2}{\varepsilon-\varepsilon_d}$. 
It can be shown that at that limit, the Newns-Anderson model translates to an effective two-level problem where one level corresponds to the adsorbate and the other corresponds to the d-band center ($\varepsilon_{\mathrm{d}}$) of the metal. In general, the poles of the chemisorption energy correspond to the energy levels of the electronic states of the adsorbate and the substrate. If we consider the poles of the Eq.\ref{sol1}, and the use the values of $\Delta$ and $\Lambda\left(\varepsilon\right)$ stated above, we get,
\begin{equation}
\varepsilon_{\pm}=\frac{\varepsilon_a+\varepsilon_d}{2} \pm \frac{1}{2} \sqrt{\left(\varepsilon_a-\varepsilon_d\right)^2+4 V^2}
\label{tow-level}
\end{equation}
The energy levels $\varepsilon_{\mp}$ describe two distinct energy levels of the metal-adsorbate system which are similar to the bonding and anti-bonding levels of H$_2$ molecule.This is the scenario of the d-band center model, however, there are differences: Newns-Anderson model does not include an explicit orthogonalization term to account for the overlap between the metal and adsorbate states ($<a|k>$), while in the d-band center model  an orthogonalization is included, something which was also considered earlier by Gimeley \textit{et. al} \cite{Grim}. Therefore, if we consider an adsorbate with only a single energy level then, we get the simple picture using which the d-band center model is derived: A picture similar to a diatomic molecule with one state per atom.  These states form bonding and anti-bonding orbitals in a manner similar to H$_2$ molecule and can be written in the form of a simple Hamiltonian given by $H=H_a+H_d+H_{ad}$, where the first and the second terms represent the adsorbate, the metal surface, and the third term representing the coupling between the two sub-systems. With a simple Linear Combination of Atomic Orbital (LCAO) type solution: $\Psi=C_a\Psi_a+C_d\Psi_d$ one gets usual solutions,
\begin{equation}
\varepsilon_{\pm}=\frac{\varepsilon_a+\varepsilon_d}{2} \pm \frac{1}{2} \sqrt{\left(\varepsilon_a-\varepsilon_d\right)^2+4 V^2}-VS
\label{tow-level}
\end{equation}
Where  $S=<\Psi_d|\Psi_a>$ is the overlap mentioned in the above. Thus 
$\varepsilon_{\pm}^{\text{d-band center}}=\varepsilon_{\pm}^{\text{Newns-Anderson}}-VS$. Thus, the Newns-Anderson model assumes that the adsorbate state is localized and does not interact with other states on the metal surface. However, for certain adsorbates, this assumption breaks down because their electronic states can interact strongly with the metal surface states and the d-band center model therefore should be more appropriate.
The chemisorption energy can be written as (within the d-band center model),
\begin{equation}
\Delta E=\underbrace{(2\varepsilon_{-}+2 f \varepsilon_{+})}_{\text{Energy after}}-\underbrace{(2 f \varepsilon_d+2 \varepsilon_a)}_{\text{Energy before}}=-2(1-f) \frac{V^2}{\left|\varepsilon_a-\varepsilon_d\right|}+2(1+f) \alpha V^2
\label{d-energy}
\end{equation}
Here $\alpha=\frac{<\Psi_d|\Psi_a>}{<\Psi_d|H_{ad}|\Psi_a>}=-\frac{S}{V}$ and  $f$ is the fractional occupation of the state $\varepsilon_d$. The above expression is derived from the basis that the adsorbate level has two electrons. Therefore, one can  obtain  the chemisorption energy in the case when the adsorbate level is empty,
\begin{equation}
\Delta E=-2f\frac{V^{2}}{|\varepsilon_{d}-\varepsilon_{a}|}+2\alpha f V^{2}
\end{equation}

According to this model, the variation in the
adsorption energy from one TM surface to another correlates with the upward shift of this d-band center ($\varepsilon_d$) with respect to the Fermi energy. A stronger upward shift indicates the possibility of the formation of a larger number of empty anti-bonding states, leading to stronger binding energy. The upward shift of the d-band center can therefore be treated as a descriptor of the catalysis. Hammer-N{\o}rskov model successfully explains both the experimental and the first-principles theoretical results for different ligands/molecules on various TM surfaces.
If we look at the above, the central prediction of the d-band center model is : 
\textit{There should be a uniform decrease (increase) of the adsorption energy of a given molecule from one TM surface to another where the number of d-electrons increases (decreases)}. Unless of course, the contribution due to the repulsive orthogonalization term does not behave irregularly, and that is not the case when one moves along a group in the periodic table.
Therefore, if we look at the adsorption energy of a given molecule on 3d-transition metals say, from Ti to Zn, the adsorption energy should change linearly as the d-occupation increases linearly. But this was not what was observed by Bhattacharjee \textit{et. al} when they looked at the adsorption energy of NH$_3$ molecule on the transition 3d-transition metal surfaces~\cite{d-band-Bhattacharjee}. Indeed there is no such \textit{linear trend} if we do a spin-polarized calculation using standard density functional theory. The adsorption energy rather behaves non-linearly.  This motivated Bhattacharjee \textit{et. al} to develop an improved version of the d-band center model where they treated the minority and majority spin band electrons separately.  By expanding the solution of the two-level problem in a spin-dependent minimum basis  $\Psi=C_{d\sigma}\Psi_{d\sigma}+C_{a\sigma}\Psi_{a\sigma}$ ($\sigma=\uparrow,\downarrow$), where the wavefunctions $\Psi_{d\sigma}$ and $\Psi_{a\sigma}$ satisfy respectively $H_d\Psi_{d\sigma}=\varepsilon_{d\sigma}\Psi_{d\sigma}$ , $H_a\Psi_{a\sigma}=\varepsilon_{a\sigma}\Psi_{a\sigma}$ and $H_{ad}\Psi_{d\sigma}=V\Psi_{a\sigma}$.\\
For a given adsorbate level $\varepsilon_a$ we get (supposing that the adsorbate is spin-unpolarized) \\
\textbf{Bonding orbitals:}\\
Up spin: $\varepsilon_{b\uparrow} = \frac{\varepsilon_{d\uparrow} + \varepsilon_a}{2} - VS - \frac{\sqrt{4V^2 + (\varepsilon_{d\uparrow} - \varepsilon_a)^2}}{2}$\\
Down spin: $\varepsilon_{b\downarrow} =  \frac{\varepsilon_{d\downarrow} +\varepsilon_a}{2} - VS - \frac{\sqrt{4V^2 + (\varepsilon_{d\downarrow} - \varepsilon_a)^2}}{2}$\\
\textbf{Antibonding orbitals:}\\
Up spin: $\varepsilon_{ab\uparrow} =  \frac{\varepsilon_{d\uparrow} +\varepsilon_a}{2} - VS + \frac{\sqrt{4V^2 + (\varepsilon_{d\uparrow} - \varepsilon_a)^2}}{2}$\\
Down spin: $\varepsilon_{ab\downarrow} =  \frac{\varepsilon_{d\downarrow} +\varepsilon_a}{2} - VS + \frac{\sqrt{4V^2 + (\varepsilon_{d\downarrow} - \varepsilon_a)^2}}{2}$\\
Thus, if we consider that  $\varepsilon_{d\uparrow}$  is lower in energy compared to $\varepsilon_{d\downarrow}$, then the energy gain due to the formation of bonding orbitals and the repulsive energy due to the formation of antibonding orbitals can be different for up spin and down spin. 
The authors considered a general situation where adsorbate has $N$ unoccupied orbitals and $M$ occupied orbitals. The chemisorption energy was obtained using a similar procedure was done for the spin unpolarized case and can be expressed as ~\cite{d-band-Bhattacharjee},
\begin{equation}
\Delta E=-\sum_{\sigma}\sum_{i=1}^{N}{\frac{f_\sigma V_{\mathrm{i}\sigma}^{2}}{|\varepsilon_{\mathrm{ai}\sigma}-\varepsilon_{\mathrm{d}\sigma}|}}-\sum_\sigma \sum_j^{M}(1-f_{\sigma}){\frac{V_{j\sigma}^{2}}{|\varepsilon_{d\sigma}-\varepsilon_{\mathrm{aj}\sigma}|}}+\sum_{\sigma}\sum_{i=1}^{N} f_{\sigma}\alpha_{i\sigma} V_{\mathrm{i}\sigma}^{2}+\sum_{\sigma}\sum_{j=1}^M (1+f_{\sigma})\alpha_{j\sigma}V_{j\sigma}^{2}
\label{ads-BWL}
\end{equation}
The improved d-band center model proposed by Bhattacharjee \textit{et. al} thus does not differ from the original Hammer-N{\o}rskov model in terms of the repulsive orthogonalization terms. In both cases, they depend on the total fractional occupation $f=\sum_\sigma f_\sigma$. But there is a rearrangement in the first two terms in Eq.\ref{ads-BWL} which are hybridization terms. It should be also noted here that the difference in the chemisorption energy between the Hammer-N{\o}rskov model and the model proposed by the Bhattacharjee \textit{et. al} would be more significant if the number of adsorbate orbitals (N\& M) are large in number.
\par
\begin{figure}
\includegraphics[scale=0.65]{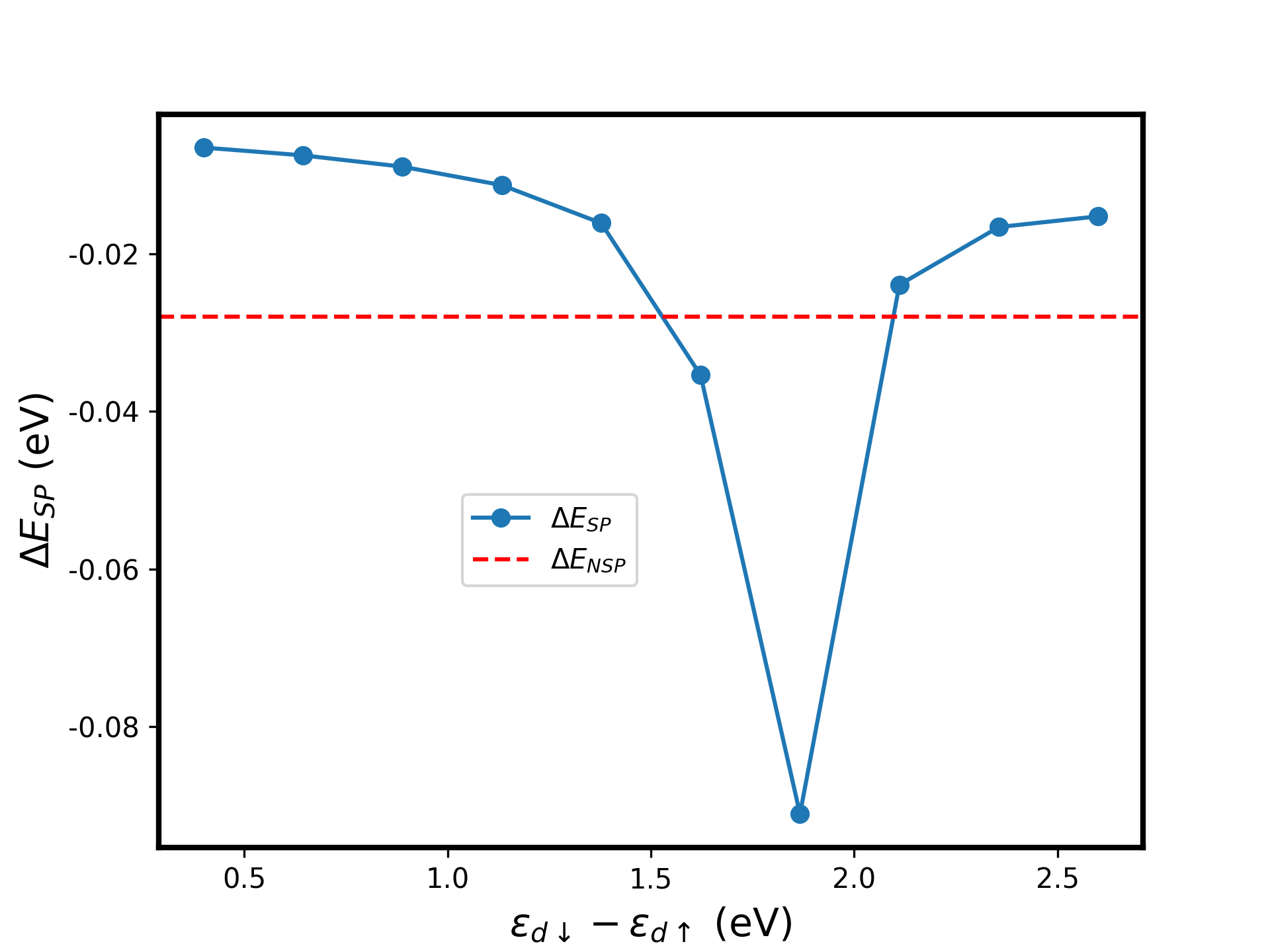}  
 \caption{Comparison of adsorption energies for an adsorbate interacting with a non-spin polarized surface ($\Delta E_{NSP}$) and spin-polarized surface ($\Delta E_{SP}$, blue line), both the surfaces have the same total fractional occupations.}
 \label{Fig:sp-nsp}
\end{figure}

To understand the difference between the above model and the standard d-band center model, let us consider a hypothetical situation: Consider a metal surface with fractional occupation $f=0.8$, interacting with an adsorbate with two orbitals, one filled $\varepsilon_1$ and other is unoccupied $\varepsilon_2$. Let us also consider a very simplistic case where adsorbate states are completely orthogonal to the d-states,i.e $\alpha=0$. From the standard d-band center model, we obtain the chemisorption energy as $\Delta E_{NSP}=-\frac{2fV^{2}}{|\varepsilon_{d}-\varepsilon_{2}|}-2{\left(1\,-\,f\right)}\frac{V^{2}}{\left|\varepsilon_{d}-\varepsilon_{1}\right|}$. Now, let us consider the effect of spin-polarization on the same surface. Let us consider now the same adsorbate interacts with the same metal surface but in a magnetically polarized state. The chemisorption energy in this case will be, $\Delta E_{SP}=-V^{2}\left[ \dfrac{f_{\uparrow }}{\left| \varepsilon_{d\uparrow} -\varepsilon_{2}\right| }+\dfrac{f_{\downarrow }}{\left| \varepsilon _{d\downarrow }-\varepsilon _{2}\right| }\right] -\left[ \left( 1-f_{\uparrow}\right) \dfrac{V^{2}}{\left| \varepsilon _{d\uparrow }-\varepsilon _{1}\right| }+\left( 1-f_{\downarrow}\right) \dfrac{V^{2}}{\left| \varepsilon _{d\downarrow }-\varepsilon _{1}\right| }\right] $. 
If we consider the following parameters: $V = 0.1 $ eV, $e_1 = -3.0 $ eV, $e_2= -2.7$ eV, $e_d =-3.5$ eV. For the spin-polarized case, we can consider $f_{\uparrow}=0.6$ and $f_{\downarrow}=0.2$, $V_\uparrow=V_\downarrow=V$. We vary the d-band centers from $-4.4$ to $0$ eV for $\varepsilon _{d\uparrow }$ and $-1.4$ to $0.4$ eV
for the $\varepsilon _{d\downarrow }$, we can see the results shown in the Fig.\ref{Fig:sp-nsp} 
We can see that in general $\Delta E_{SP}$ more positive than $\Delta E_{NSP}$ exceptional cases are when either of $\varepsilon _{d\uparrow}$ or $\varepsilon _{d\downarrow }$ are very close to the adsorbate levels~($\varepsilon _{1}$ or $\varepsilon _{2}$). 
It was demonstrated by Bhattacharjee \textit{et al.} that in the case of NH$_3$ adsorption on the 3d transition metal surfaces, the adsorption energy calculated using the spin-generalized d-band center model works better than the standard Hammer-Norskov approach for the ferromagnetic surfaces like Fe, Co, Mn, etc while for the unpolarized surfaces both the models give the same results. Bhattacharjee \textit{et al.} also proposed a descriptor that can be used instead of the standard d-band center model, which is an occupation-weighted sum of the d-band centers for the majority and minority spin electrons and is associated with a shift proportional to the exchange split of metal and is given by,
\begin{equation}
\epsilon^{eff}=\sum_{\sigma}\frac{f_{\sigma}\varepsilon_{d\sigma}}{\sum_{\sigma}f_\sigma}-\left(\varepsilon_{d_{\downarrow}}-\varepsilon_{d\uparrow}\right)\mu
\label{effective}
\end{equation}
Here $\mu=\frac{f_{\uparrow}-f_{\downarrow}}{f_{\uparrow}+f_{\downarrow}}$ is the reduced fractional occupations. It is interesting to note that the above descriptor was recently used to formulate the effective energy barrier for ion migration in halide perovskites~\cite{migration} beyond the usual catalytic applications.
\par
The  \textit{spin-corrected} d-band center model offers an insightful explanation for the observed disparity in hydrogen-metal binding strength between ferromagnetic and antiferromagnetic Fe surfaces~\cite{melander2014effect}. In this framework, the chemical reactivity of antiferromagnetic materials is elucidated through the lens of the antiferromagnetic-to-ferromagnetic transition and the spin sub-lattices A and B. The authors point out that in antiferromagnetic systems, two distinct spin sub-lattices, labeled as A and B, exist. When both sub-lattices consist of the same metal, the energy levels of the d-orbitals within these sub-lattices exhibit opposite signs. Consequently, this asymmetry in the interaction between the metal and adsorbates affects the strength of hydrogen-metal bonds in the majority and minority spin channels. As a result, hydrogen-metal binding is weaker on a ferromagnetic Fe surface, where only minority spin electrons participate in bond formation, in contrast to an antiferromagnetic Fe surface, where both minority and majority spin electrons contribute to bond formation~\cite{d-band-Bhattacharjee}.

It is important to note that the \textit{spin-corrected} d-band center model discussed above remains suitable for catalysts characterized by a relatively weak electron-electron correlation, or more precisely, for catalysts whose electronic structure can be well described by Density Functional Theory (DFT). This applicability extends to both the original and spin-corrected versions of the d-band center model. These models serve as valuable tools for guiding the determination of adsorption energies through standard DFT calculations, as expressed by the equation,
\begin{equation}
E_{ads}=E_{S+A}-(E_S+E_A)
\end{equation}
where $E_{S+A}$, $E_S$, and $E_A$ are respectively energy of the surface with adsorbate, the energy of the surface, and the energy of the adsorbate alone. These energies are obtained with the standard DFT-based approaches. For example, for the surface, the total energy is given by
\begin{equation}
E[\rho_S]=\int \rho_S(\boldsymbol{r}) v_{\text{ext}}(\boldsymbol{r}) \mathrm{d} \boldsymbol{r}+T[\rho_S]+U[\rho_S]+E_{\mathrm{XC}}[\rho_S]=\int \rho_S(\boldsymbol{r}) v_{\text{ext}}(\boldsymbol{r})+F[\rho_S]
\end{equation}
Here, the universal functional $F[\rho_S]$ is a sum of the electronic kinetic energy $\mathrm{T}[\rho_S]$, classical Coulomb repulsion energy $\mathrm{U}[\rho_S]$, and quantum exchange-correlation energy $E_{X C}[\rho_S]$ functionals.
$\rho_S$ being the electron density and $v_{ext}$ is the external potential. For the spin-polarized system, the total energy is
\begin{equation}
E[\rho_S^{\uparrow},\rho_S^{\downarrow}]=\int \rho_S(\boldsymbol{r}) v_{\text{ext}}(\boldsymbol{r})+F[\rho_S^{\uparrow},\rho_S^{\downarrow}]
\end{equation}
The mathematical expression for chemisorption energy $\Delta E (f,V,\varepsilon_{d\sigma})$, derived from Eq.\ref{Newns-Anderson} and expressed in Eq.\ref{ads-BWL}, lacks an explicit representation of the correlation effects. This omission arises from the absence of an electron-electron correlation term in $H_d$ of the model Hamiltonian, $H=H_d+H_a+H_{ad}$, (By inclusion of correlation effect, $H_d$ would be, $H_d=\sum_k \epsilon_k c_k^{\dagger} c_k+\frac{U}{N} \sum_{k, p, q} c_{p+q \uparrow}^{\dagger} c_{k-q \downarrow}^{\dagger} c_{p \uparrow} c_{k \downarrow}$, where $U$ represents the on-site Coulomb energy term usual for the Hubbard model, and $N$ is the number of sites. Within the above d-band center models, the correlation effects manifest indirectly through $\varepsilon_{d\sigma}$, which is usually determined via density functional theory-based calculations and used as input for the calculation of chemisorption energy $\Delta E$. These calculations incorporate exchange and correlation effects through $E_{\mathrm{XC}}[\rho]$ as mentioned above.

However, for the transition metal oxide catalysts usually used in OER and ORR, one may need to consider the effect of the strong electron correlation~c\cite{tran2023open}. For example, recent works by Gracia \textit{et al} discuss that strongly correlated electrons play a crucial role in the energy of magnetic materials and thus in catalysis~\cite{gracia2017spin,biz2022review,biz2021strongly}. While catalysis involves a local picture of atoms forming bonds on the surface with one or a few atoms, the electronic structure of the surface can be strongly influenced by the presence of unpaired electrons and their interactions. Gracia \textit{et al} propose that these interactions can lead to cooperative quantum spin exchange interactions (QSEI) and quantum excitation interactions (QEXI), which can affect the efficiency, cost, implementation, and environmental footprint of any heterogeneous catalyst~\cite{biz2021strongly}. Understanding the role of strongly correlated electrons in catalysis is therefore crucial for the design of more efficient and selective catalysts.
Furthermore, such oxides often exhibit complex electronic structures with localized electron configurations and strong electron correlation effects. The binding energies of adsorbed oxygen species on these oxides can be significantly influenced by electron correlation. For example, electron correlation can lead to the formation of localized electronic states within the bandgap, affecting the charge transfer between adsorbates and the oxide surface. This, in turn, impacts the stability and reactivity of adsorbed species.
\par
Even for metallic systems, it may be necessary to go beyond the one-electron picture, as suggested by the recent work of Jans \textit{et al}~\cite{janas2023enhancing}, in which the authors discuss how the adsorption of oxygen atoms on the surface of ferromagnetic iron leads to a breakdown of the Stoner-like picture of band ferromagnetism. This is evidenced by the reduction of the exchange splitting, the narrowing and localization of hybrid O-Fe states, and the appearance of spin-polarized adsorbate-FM satellite features.
\par
While the above study reports that chemisorption affects band ferromagnetism by collapsing the exchange splitting close to the Fermi level, there are other studies that report the opposite behavior. In certain 3d transition metals, chemisorption may affect the Stoner criterion for the appearance of magnetism, where the interface between metallic thin films and C60 molecular layers modifies the states of metal in such a way that they can overcome the Stoner criterion, resulting in the emergence of ferromagnetic behavior at room temperature in non-ferromagnetic materials~\cite{beating}. Chemisorption can modify the density of states (DOS) at the Fermi energy, while the Stoner criterion depends on the DOS as well. By considering the coupling between the adsorbate states and the metal states, a modified Stoner criterion was proposed by Bhattacharjee \textit{et al} to analyze how chemisorption may overcome the barrier laid by the standard Stoner criterion and lead to the appearance of ferromagnetism on otherwise non-magnetic metal surfaces~\cite{stoner}. The authors argue that in the presence of chemisorption, the standard Stoner criterion for ferromagnetism is $D(E_F)I\ge 1$ (where $D(E_F)$ is the density of states (DOS) at the Fermi energy and $I $ is the stoner parameter) should be changed to $D(E_F)I\ge 1-I\lambda $~\cite{stoner}. Here the factor $\lambda=D_a(E_F)[1+\frac{1}{\pi}\left(Im\frac{d\Sigma(E)}{dE}\right)_{E_F}]$ considered to be greater than zero to allow a nonmagnetic surface to become ferromagnetic, which otherwise fails to satisfy Stoner's criterion. $D_a(E_F)$ is the DOS of the adsorbate at the Fermi energy and $\Sigma(E)$ describes the self-energy due to the interaction between the adsorbate and the metal surface.

\par
Another essential element of the magnetism reliance on the reactivity of the catalysts can be visualized using the magnetic field, particularly in oxygen-based electrochemical processes such as ORR and OER~\cite{field1,field2,field3}. Some recent investigations have reported the dependence of the ORR current on the external magnetic field. However, these studies are based on outer shell mechanisms, such as dependence on oxygen ion concentration, magnetic field gradient, and so on, and do not address how magnetic field interferes with the chemical binding of oxygen reduction reaction intermediates. In ORR we obtain non-magnetic water from the paramagnetic oxygen, while in OER we can obtain paramagnetic oxygen from the non-magnetic water. 
In recent work, Bhattacharjee \text{et al.} used constrained density functional theory formalism to investigate the role of spin orientation on the reactivity of ORR intermediates  on a ferromagnetic electrode surface~\cite{Controlling}. They considered the direct dissociative process,
\begin{equation}
\begin{gathered}
\frac{1}{2} \mathrm{O}_2+* \rightarrow \mathrm{O}^* \\
\mathrm{O}^*+\left(\mathrm{H}^{+}+\mathrm{e}^{-}\right) \rightarrow \mathrm{HO}^* \\
\mathrm{HO}^*+\left(\mathrm{H}^{+}+\mathrm{e}^{-}\right) \rightarrow \mathrm{H}_2 \mathrm{O}+*
\end{gathered}
\end{equation}

\begin{figure}
\includegraphics[scale=0.65]{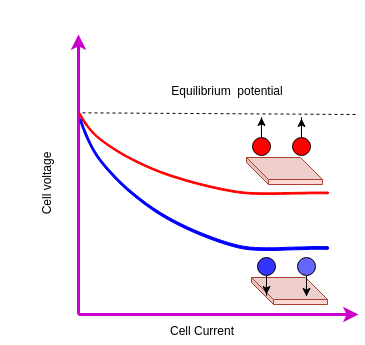}  
 \caption{Schematic showing the behavior of the overpotential in the ORR reaction. The blue curve corresponds to the case when oxygen atoms spins are unconstrained, while the red curve corresponds to the case when the oxygen atoms spins are constrained.}
 \label{Fig:Schem}
\end{figure}

Where "*" represents the adsorption site, and they showed that the strength of binding for these reaction intermediates (O*, HO*) depends on their relative spin orientations with respect to the magnetization of the electrode. The authors suggest that oxygen-based electrochemical reactions can be controlled through an applied magnetic field and demonstrate this possibility by studying an ORR on a PdFe (001) surface using a new concept called \textit{spin orientation dependent overpotential}. The adsorption energy was calculated using: $E_{ads}^{\Uparrow,\uparrow(\downarrow)}=E_{S\Uparrow+O \uparrow(\downarrow)}-\left(E_{S}\Uparrow+\frac{1}{2} E_{O_{2}}^{(g)}\right)$.
Where $\Uparrow$ represents the electrode spin moment and $\uparrow(\downarrow)$ is the direction of the spin moment of the oxygen species. To fix the magnetic moment of the reaction intermediates they used penalty energy given by, $E\left[\rho,\left\{\hat{\mathbf{M}}_{\mathbf{I}}^{\mathbf{d}}\right\}\right]=E_{D F T}[\rho]+E_{p}=E_{D F T}[\rho]+\sum_{I} \lambda_{I}\left(\left|\mathbf{M}_{I}\right|-\hat{\mathbf{M}}_{\mathbf{I}}^{\mathrm{d}} \cdot \mathbf{M}_{\mathbf{I}}\right)$
The penalty energy term in the above equation is given by $E_{p}=\sum_{I} \lambda_{I}\left(\left|\mathbf{M}_{I}\right|-\hat{\mathbf{M}}_{\mathrm{I}}^{\mathrm{d}} \cdot \mathbf{M}_{\mathrm{I}}\right)$. $\hat{\mathbf{M}}_{\mathrm{I}}^{\mathrm{d}}$ is a unit vector along the desired direction of the magnetic moment and $\rho$ is the electron density. The overpotential was calculated by from Gibb's free energy $\Delta G=\Delta E+\Delta Z P E-T \Delta S$.  Without any constraint, the spin moments of the reaction intermediates (O* and HO*) are anti-parallel to the electrode spin-moment. They obtained two different overpotentials from Gibb's free energy for the oxygen reduction reaction on a PdFe (001) surface. The overpotentials are respectively 0.43 V and  0.2 V, for the spin-constrained case, where they align the spin moments of the oxygen intermediates parallel to the electrode moment. Thus the overpotential is reduced by about half. The schematic in Fig.\ref{Fig:Schem} illustrates the situation. Here, only the oxygen atoms have indicated arrow directions. Upward arrows signify that the spins of the oxygen atoms align with the surface spin moments (which aren't depicted in the figure). Conversely, downward arrows denote that the spins of the oxygen atoms are opposite to the surface magnetic moments.
In their study, the authors provide an outlook on the potential applications of their findings. They suggest that the concept of spin orientation could be used to design new catalysts with improved electrochemical performance. Additionally, the results of their study could have implications for the development of more efficient fuel cells and other electrochemical devices. Further research in this area could lead to exciting new advances in the field of electrochemistry.
\par

Spin-dependent interactions, such as exchange interactions, spin-orbit coupling, and magneto-electric effects, play a crucial role in oxygen-based electrochemical reactions such as OER and ORR. Gracia \textit{et al } pointed out~\cite{gracia2017spin} that during OER and ORR, where the number of unpaired electrons is not conserved, magnetic potentials in optimal catalysts act as selective gates, enhancing the transport of local spin currents and facilitating electron transfer over the conduction band. Spin-dependent forces decrease the activation energy by decreasing the bonding properties of donor or acceptor orbitals in the catalyst, allowing for easier electron transfer.
\par
The concluding theme we wish to discuss is a significant and intriguing concept in heterogeneous catalysis: The scaling relationships, which pose a constraint in the engineering of catalysts for crucial reactions such as Oxygen evolution reaction (OER) catalysts~\cite{jones2008using,craig2019universal,omojola2022site,zhao2019theory,calle2012physical}. The scaling relation of adsorption energies usually refers to the linear relationships between the adsorption energies of different reaction intermediates on a catalyst surface~\cite{abild2007scaling}. These relationships are important in heterogeneous catalysis because they provide insights into the reactivity and selectivity of catalytic reactions. While this can simplify the computational modeling of catalytic reactions, it can also pose a challenge for the design of catalysts with enhanced activities, as linear dependence may preclude further optimization of catalytic activities. In other words, the well-defined energy connection between adsorbates is said to be detrimental to catalysis. The most intriguing aspect of this scaling relationship is that the slope of the linear scaling relationship between the adsorption energies of two reaction intermediates is determined only by their valency~\cite{abild2007scaling}. The surface variables do not appear in the mathematical description of the slope. The original scaling relationship~\cite{abild2007scaling} was put forward for the adsorption energies for the hydrogen containing molecules AH$_x$ (A=C, O, H, N, and S) on different metal surfaces, denoted as $E_{\text{ads}}(AH_x)$, and the adsorption energy of the central atom A, denoted as $E_{\text{ads}}(A)$, is given by:
\begin{equation}
E_{\text{ads}}(AH_x) = \gamma E_{\text{ads}}(A) + \xi
\end{equation}
where  $E_{\text{ads}}(AH_x)$ is the adsorption energy of the species $AH_x$, $E_{\text{ads}}(A)$ is the adsorption energy of the species (A), $\gamma$ is the scaling factor or slope, and $\xi$ is the intercept. 
The scaling relationship is derived from the concept of effective medium theory and the d-band center model: (1) Effective medium theory predicts an optimal electron density $n_0$ that results in the lowest energy position for an atom~\cite{abild2007scaling}. An adsorbing atom obtains electron density from all the atoms it is bonded to, including other atoms in the adsorbate as well as surface atoms. Assuming that other adsorbate atoms all contribute a density of $n_0\left(\frac{x}{x_{max}}\right)$, the density contributed by the surface can be calculated and is given by, $n_{sur}=n_0(\frac{x_{max}-x}{x_{max}})$. Here $x_{max}$ is the maximum number of H atoms that can form stable bonds to the A atom. Considering the adsorption energy has a contribution from both $sp$-electrons and $d$-electrons, i.e $E_{\text{ads}}=\Delta E_{d}+\Delta E_d$, it was assumed that  $E_d(AH_x)=\gamma(x)\Delta E_d(A)$. This means that the slope depends on the adsorbates coupling to the d-states alone. (2) According to the d-band center model, the change in adsorption energy ($\Delta E_d$) is proportional to ($V_{ad}^2$) since the coupling between the adsorbate and the surface scales approximately with density ($n_{surf} $). Therefore ($\Delta E_d \propto V_{ad}^2 \propto n_{surf} \propto (x_{max} - x)/x_{max}$)~\cite{montemore2014scaling}. The adsorption energy scaling theory is based on the effective medium theory, which is further based on a homogeneous description of the electron gas. This assumption may not hold true for all systems, particularly for highly spin-polarized surfaces and nanostructures with inhomogeneous electron density. It should be noted here that according to the effective medium theory, there is no change to exchange and correlation energies upon the addition of the \textit{ad-atoms}. As a result, depending on a homogenous description of electron gas (such as the jellium model) to guide the results obtained via methodologies such as density functional theory may not always be effective.
\par
The task of developing efficient electrocatalysts using non-precious metals is a complex challenge, as optimal catalytic materials necessitate a balance in bonding with key intermediates~\cite{sabatier1913catalyse,balandin1969modern}. Therefore, enhanced comprehension of quantitative structure-activity relationships is vital for the creation of innovative and highly efficient catalysts~\cite{hong2016descriptors,li2022oxygen,cao2018mechanistic}.

While magnetic bimetallic transition metals (TMs) constitute an intriguing category of materials, their involvement in scaling relationships has been relatively explored~\cite{khorshidi2018strain,wang2017breaking,calle2017covalence}. Most TM compounds researched in the academic literature are non-magnetic, hence the impact of surface magnetization on adsorption energy and scaling laws is not usually considered~\cite{chretien20082,mtangi2015role}. Recent investigations suggest that magnetic bimetallic TMs may disrupt scaling relationships, paving the way for developing innovative and superiorly efficient catalysts~\cite{vojvodic2015new,khorshidi2018strain}. Recent works by Ram \textit{et al}~\cite{ram2020adsorption,ram2021identifying} elaborated on the significance of surface descriptors within the context of scaling relationships, notably demonstrating the dissolution of standard scaling laws for hydrogenated species of oxygen (O), carbon (C), and nitrogen (N) on magnetic bimetallic TM surfaces such as NiPt, FePt, FePd, MnPt, MnPd, CoPt, CoPt$_3$, MnPt$_3$, FePt$_3$, NiPt$_3$, and Co$_3$Pt. Additionally, they highlighted that so-called negative scaling relationships between adsorption energies could be derived using surface descriptors like magnetic moments and d-band centers.

As already mentioned, the usual scaling relationship~\cite{abild2007scaling} states that the chemisorption energy, $\Delta E_{ads}$, between two adsorbates 1 and 2 on transition metal surfaces are related by,

\begin{equation}
\Delta E_{ads}^1 = \gamma \Delta E_{ads}^2 +\xi
\label{equ:ads}
\end{equation}
Here, $\Delta E_{ads}^1$ and $\Delta E_{ads}^2$ denote the respective adsorption energies of species 1 and 2. It is postulated that on a densely packed surface, the slope, $\gamma$, relies on the ratio of the valence electrons' number of the adatoms attached to the surface,\cite{abild2016computational} and that the intercept, $\xi$, depends on the surface coordination.\cite{calle2015introducing} Consider a set of surface variables ({$\omega_i$}) influencing the $\Delta E_{ads}$ of species 1 and 2, which subsequently correlate the $\Delta E_{ads}^1$, $\Delta E_{ads}^2$ of species 1 and 2, respectively, as

\begin{equation}
\Delta E_{ads}^{1} = F(\lbrace\omega_i\rbrace) + \alpha_{0}\quad and \quad
\Delta E_{ads}^{2} = G(\lbrace\omega_i\rbrace) + \beta_{0}
\label{equ:function}
\end{equation}

Further, equation-\ref{equ:ads} on magnetic bimetallic TMs must obey the following constraints, expressed by equation-\ref{equ:offset} below \cite{calle2012physical,calle2015introducing}

\begin{equation}
F (\lbrace\omega_i\rbrace) = \gamma G (\lbrace\omega_i\rbrace) \quad and \quad
\xi = \alpha_0 -\gamma\beta_0
\label{equ:offset}
\end{equation}

\begin{figure}
  \centering
     \subfigure[]{\includegraphics[scale=1.3]{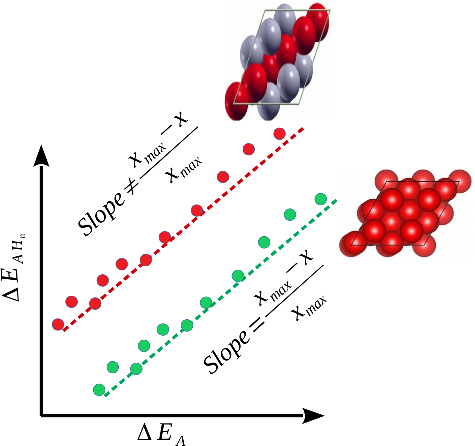}}  
     \subfigure[]{\includegraphics[scale=1.3]{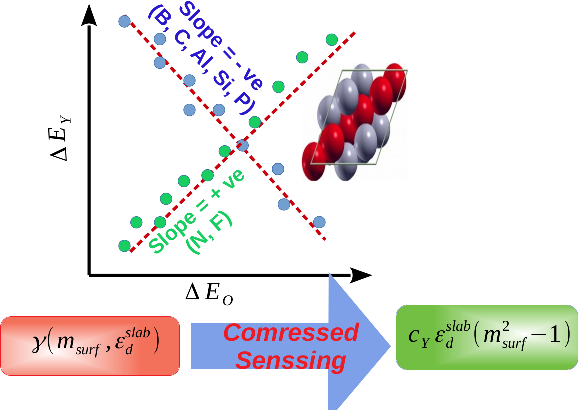}}
 \caption{ Figure  presents a schematic diagram illustrating the adsorption energy scaling relation on bimetallic magnetic surfaces. The standard scaling relationship breakdown on bimetallic surfaces is depicted in Figure \ref{Fig:scaling} (a). Meanwhile, Figure \ref{Fig:scaling} (b) explains the negative scaling relationship observed between O and Y (where Y = N, F, B, C, Al, Si, P). In both scenarios, the slope appears to be influenced by the surface descriptors.}
 \label{Fig:scaling}
\end{figure}

The predicted slope value (0.5) by the N{\o}rskov group\cite{abild2007scaling} and that of Ram \textit{et al} \cite{ram2020adsorption} between O* and its hydrogenated species on monometallic non-magnetic transition metal surfaces for the scaling relation align remarkably well. However, for bimetallic magnetic TM surfaces, the slope deviates from 0.5.\cite{ram2020adsorption} For these magnetic bimetallic surfaces, the scaling relation's slope also diverges from expected values for other hydrogenated C and N species\cite{abild2007scaling} (refer to Figure-\ref{Fig:scaling} (a)). To substantiate their claim, the authors showed  that all combinations of the energy scaling relation for adsorption comply with equation-\ref{equ:offset}, considering the specific descriptor, which in these instances is the surface's magnetic moment. Magnetic bimetallic TM surfaces demonstrate a similar linear correlation between the adsorption energies of the most stable site for adsorbed O and other atomic adsorbates (B, C, N, F, Al, Si, and P) from the second and third rows of the periodic table.\cite{ram2021identifying}
This research exhibits a positive slope for O versus N and F, but a negative slope for O versus other specified atomic adsorbates including B, C, P, Al, and Si on the magnetic bimetallic TMs.
Scaling relationships and negative slopes among atomic adsorbates have already been observed for various non-magnetic surfaces.
\cite{calle2012physical,su2016establishing}

The potential to customize the geometry and composition of surface atoms for desired properties critically depends on the ability to differentiate between simple descriptors and complex electronic properties of materials. The authors propose that surface properties such as the average number of valence electrons ($NV_{av}^{slab}$), the surface's magnetic moment ($m_{surf}$), the work function ($\phi_{slab}$), and the d-band center of the surface ($\epsilon_d^{slab}$) as key parameters in determining the adsorption energy scaling relation. Independent variables can be chosen using Pearson's correlation, and equation-\ref{equ:ads} can be rephrased as

\begin{equation}
\Delta E_{ads}^{1} = \gamma (m_{surf}, \epsilon_d^{slab}, \phi_{slab}, NV_{av}^{slab} ) \Delta E_{ads}^{2} +\xi \
\label{equ:scaling2}
\end{equation}
instead,
$\Delta E_{ads}^{1} = \gamma (x_{max}, x) \Delta E_{ads}^{2} +\xi $ as originally reported by Abild-Peterson \textit{et. al} \cite{abild2007scaling} see Figure-\ref{Fig:scaling}(a))

The magnetic moment is pivotal in estimating adsorption energy and, consequently, in the linear scaling relation for atomic adsorbates and their hydrogenated species on bimetallic surfaces.\cite{ram2020adsorption} The $\epsilon_d^{slab}$ and $m_{surf}$ are identified as two independent descriptors in describing the scaling relation among atomic adsorbates on bimetallic TM surfaces.\cite{ram2021identifying}
The authors utilized the sure-independence screening and sparsifying operator (SISSO) method~\cite{SISSO} to derive an analytical form of $\gamma$ ($\epsilon_d^{slab}$, $m_{surf}$). SISSO is a compressed-sensing method that can perform feature engineering and identify the best low-dimensional descriptor in a large set of candidates. SISSO can generate new features by combining original features in various ways. It ensures that the selected features are Surely independent of each other, avoiding redundancy. Finally, it aims to find a sparse model, meaning it tries to explain the data with as few features as possible.
Considering the ratio $\left(\frac{\Delta E_Y}{\Delta E_O}\right)$ as the target (where Y = N, F, B, C, Al, Si, P) and performing a multi-tasking learning with optimization of, $\underset{C}{\operatorname{argmin}} \sum_{k=1}^{7} \frac{1}{N_k^M}||\left(\frac{\Delta E_Y}{\Delta E_O}\right)^k-D^kC^k||^2_2+\lambda||C||_0$ a one-dimensional descriptor was selected in such a way that it has a strong correlation with the target. Here,  $D^k=(d^k_{j,l})$ is the $(M\times N)$ dimensional matrix of the inputs known as \textit{sensing matrix} and C is the coefficient vector. $N_k^M$ is the number of samples.  Here $k$ ranges from 1 to 7 as we are considering seven adsorbates on different suraces.
Based on this choice, it was determined that the scaling relationship for a given adsorbate "Y".(refer to Figure-\ref{Fig:scaling} (b))
\begin{equation}
\Delta E_Y=\Delta E_O\left[c_Y\epsilon_d^{slab}(m_{surf}^2-1)\right]+\beta_Y
\end{equation}
The intercept $C_Y$ is positive for all the elements except N and F with which $\Delta E_O$ has a negative slope. $\beta_Y$ is the usual intercept. It should be remembered that the $\epsilon_d^{slab}$ is negative and $(m_{surf}^2-1)>0$. It is interesting to note that, the functional form of the above slope obtained by the compressed sensing approach suggests that one can tune it by changing the surface descriptor such as magnetic moments.
The fact that the slope depends on surface descriptors such as magnetic moments suggests that for ferromagnetic surfaces, one must consider factors beyond the valency of the adsorbate. The standard scaling relationship suggests that if two adsorbates have the same valency, the slope $\gamma$ should be equal to one. However, this may not be valid in the case of a magnetic surface, as these two molecules could change the exchange-correlation energy of the metal surface in different ways due to chemisorption and therefore the slope could be different than one.

The original scaling relationships were demonstrated ~\cite{abild2007scaling,montemore2014scaling}, performing a simple linear regression of the adsorption energies to find the slope and intercept of the linear relationship between the adsorption energies on the reference surface and the other surfaces. As most of the surfaces were non-magnetic, the dependence  on the surface variable such as magnetic moment did not appear in such simple statistical analysis. Therefore, it is not reasonable to assume that the adsorption energies between two adsorbates would have the same slope on a non-magnetic surface and another one with a large magnetic moment. Indeed, if these two adsorbates have magnetic moments (as in the case of OER or ORR reaction intermediates such as O*, HO* or HOO* etc), there will be magnetic moment dependent surface adsorbate interaction with energy $E_{int} = -J \sum_j {\bf m}^a \cdot {\bf m}^S_j$. Here $J$ represents the exchange coupling constant, which quantifies the strength of the exchange interaction between surface moments ${\bf m}^S_j$ and the adsorbate moment ${\bf m}^a$. $j$ represents the number of magnetic atoms on the surface with which the adsorbate interacts.  
Furthermore, if the surface has a non-trivial magnetic arrangement (such as A-type or G-type AFM order), the role of surface moments could be even more prominent as pointed out by some recent studies~\cite{biz2020catalysis}.
\par
An interesting approach for future investigations lies in using the time-dependent Newns-Anderson model to study the dynamic behavior of paramagnetic molecules such as O$_2$ on metal surfaces. While previous studies have successfully used this model to study hydrogen atoms, expanding its application to paramagnetic molecules is a promising direction. previous studies have shown that nonadiabatic effects are pronounced near the spin transition point, making the near-adiabatic approximation insufficient. Remarkably, this suggests that the spin transition has a significant impact on the nonadiabatic energy transfer and system dynamics during adsorption. The spin transition, which denotes the shift of the ground state of the system from the spin-polarized to the non-polarized state, is characterized by a transformation of the solutions into spin-dependent adsorbate states, which evolve from distinct solutions to degenerate ones. By examining the time-dependent Newns-Anderson model in the context of paramagnetic molecules, we can unravel the complex interplay between spin dynamics and nonadiabatic effects and shed light on the behavior of these systems at the atomic level.
\par
Although not covered in this article, it is worth highlighting another area where 3D transition metals are being explored: single-atom catalysts~\cite{SAC1,SAC2}. Recent investigations have shown that single-atom catalysts (SACs) with significant metal loading have emerged in several heterogeneous catalytic sectors and exhibit exceptional catalytic properties. For example, single-atom 3d block transition-metal catalysts have been applied as electrocatalysts for the CO$_2$ reduction process, with Ni and Fe SACs showing improved catalytic activity for CO evolution~\cite{SAC2}. It goes without saying that there will be several magnetism-related phenomena in the catalytic reactions on such catalysts.
\section{Acknowledement}
This work was supported by the Korea Institute of Science and Technology, GKP (Global Knowledge Platform, Grant number 2V6760) project of the Ministry of Science, ICT and Future Planning.
\clearpage
\newpage
\bibliographystyle{}
\bibliography{ref}
\end{document}